\documentclass[conference]{IEEEtran}
\IEEEoverridecommandlockouts
\usepackage{cite}
\usepackage{amsmath,amssymb,amsfonts}

\usepackage{color, soul}

\usepackage{url}
\usepackage{graphicx}
\usepackage{float} 
\usepackage{subfigure}
\usepackage{algorithm}
\usepackage{algpseudocode}
\usepackage{amsmath}
\usepackage{multicol}
\usepackage{multirow}
\usepackage{algorithm,algpseudocode,float}
\usepackage{lipsum}
\makeatletter
\newenvironment{breakablealgorithm}
  {% \begin{breakablealgorithm}
   \begin{center}
     \refstepcounter{algorithm}% New algorithm
     \hrule height.8pt depth0pt \kern2pt% \@fs@pre for \@fs@ruled
     \renewcommand{\caption}[2][\relax]{% Make a new \caption
       {\raggedright\textbf{\ALG@name~\thealgorithm} ##2\par}%
       \ifx\relax##1\relax % #1 is \relax
         \addcontentsline{loa}{algorithm}{\protect\numberline{\thealgorithm}##2}%
       \else % #1 is not \relax
         \addcontentsline{loa}{algorithm}{\protect\numberline{\thealgorithm}##1}%
       \fi
       \kern2pt\hrule\kern2pt
     }
  }{% \end{breakablealgorithm}
     \kern2pt\hrule\relax% \@fs@post for \@fs@ruled
   \end{center}
  }
\makeatother

\def\BibTeX{{\rm B\kern-.05em{\sc i\kern-.025em b}\kern-.08em
    T\kern-.1667em\lower.7ex\hbox{E}\kern-.125emX}}
\begin{document}

\title{{Deep Learning-Based Strategy For Macromolecules Classification with Imbalanced Data from Cellular Electron Cryotomography}
}    

\author{\IEEEauthorblockN{1\textsuperscript{st} Ziqian Luo}
\IEEEauthorblockA{\textit{International School} \\
\textit{Beijing University of Posts and Telecommunications}\\
Beijing, China\\
luoziqian@bupt.edu.cn}
\and
\IEEEauthorblockN{2\textsuperscript{nd} Xiangrui Zeng}
\IEEEauthorblockA{\textit{Computational Biology Department} \\
\textit{Carnegie Mellon University}\\
Pittsburgh, United States \\
xiangruz@andrew.cmu.edu}
\and
\IEEEauthorblockN{3\textsuperscript{rd} Zhipeng Bao}
\IEEEauthorblockA{\textit{Department of Electronic Engineering} \\
\textit{Tsinghua University}\\
Beijing, China\\
bzp15@mails.tsinghua.edu.cn}
\and
\IEEEauthorblockN{4\textsuperscript{th} Min Xu\thanks{Corresponding author: Min Xu}}
\IEEEauthorblockA{\textit{Computational Biology Department} \\
\textit{Carnegie Mellon University}\\
Pittsburgh, United States \\
mxu1@cs.cmu.edu}
 }

\maketitle

\begin{abstract}
Deep learning model trained by imbalanced data may not work satisfactorily since it could be determined by major classes and thus may ignore the classes with small amount of data. In this paper, we apply deep learning based imbalanced data classification for the first time to cellular macromolecular complexes captured by Cryo-electron tomography (Cryo-ET). We adopt a range of strategies to cope with imbalanced data, including data sampling, bagging, boosting, Genetic Programming based method and. Particularly, inspired from Inception 3D network, we propose a multi-path CNN model combining focal loss and mixup on the Cryo-ET dataset to expand the dataset, where each path had its best performance corresponding to each type of data and let the network learn the combinations of the paths to improve the classification performance. In addition, extensive experiments have been conducted to show our proposed method is flexible enough to cope with different number of classes by adjusting the number of paths in our multi-path model. To our knowledge, this work is the first application of deep learning methods of dealing with imbalanced data to the internal tissue classification of cell macromolecular complexes, which opened up a new path for cell classification in the field of computational biology.
\end{abstract}

% \begin{IEEEkeywords}
% Deep Learning, Imbalanced classification, Convolutional Neural Networks
% \end{IEEEkeywords}

\section{Introduction}
Biological pathways depend on the function of macromolecular complexes, whose structure and spatial organization are critical to the function and dysfunction of pathways. Due to the limitations of data acquisition techniques, the native structural information of macromolecular complexes is extremely difficult to obtain \cite{b1}. With the development of biotechnology, Cryo-electron tomography (Cryo-ET) enables 3D visualization of cellular tissue in near-native state and sub-molecular resolution \cite{b2,b3,b4}, making it a powerful tool for analyzing macromolecular complexes and their spatial organization within single cells \cite{b5}. 

However, it is often observed that the macromolecular complex data collected is imbalanced. The protein concentration difference can be as large as seven orders of magnitude \cite{b6}. One type of macromolecular complexes may dominate over other types, resulting in a low accuracy. In fact, the problem of data imbalance also occurs in most real-world classification problems. The collected data follows a long tail distribution i.e., data for few object classes is abundant while data for others is scarce. This phenomenon is termed the data-imbalanced classification problem \cite{b7}. Although the problem of data-imbalanced classification occurs frequently in the computer vision field, research work on this topic has been rare in recent years. Almost all competitive datasets avoid data-imbalanced during the evaluation and training procedures. For instance, the case of the popular image classification datasets (such as CIFAR−10/100, ImageNet, Caltech−101/256, and MIT−67), efforts have been made by the collectors to ensure that, either all of the classes have a minimum representation with sufficient data, or that the experimental protocols are reshaped to use an equal number of images for all classes during the training and testing processes \cite{b8,b9}.

In this paper, we conduct extensive experiments and explore various methods for dealing with the problem of data-imbalanced classification, such as data sampling, bagging, boosting, Genetic Programming based method. We rigorously prove that the above various methods are indeed effective, and apply them to the classification of macromolecular complexes in single cells, and achieved a competitive classification performance. In particular, we make the following key contributions:
\begin{itemize}
\item We summarize various well-known methods for dealing with data-imbalanced classification problems in order to improve classification performance and further find the best combinations with our own model among the methods.

\item We apply the method of dealing with imbalanced data to the classification of cell macromolecular complexes for the first time.

\item We propose a novel method to solve the data-imbalanced problem, termed as multi-path convolutional neural network (CNN), which significantly improves the classification performance over traditional methods. Moreover, our model is flexible enough to cope with different number of classes by adjusting the number of paths in our multi-path CNN model.

\end{itemize}

\section{Related Work}

\subsection{Cryo-electron tomography (Cryo-ET)}

Cryo-electron tomography (Cryo-ET) is an imaging technique used to produce high-resolution ($4 nm$) three-dimensional views of samples, typically biological macromolecules in cells \cite{b9}. Cryo-ET is a specialized application of transmission electron cryomicroscopy (CryoTEM) in which samples are imaged as they are tilted, resulting in a series of 2D images that can be combined to produce a 3D reconstruction, similar to a CT scan of the human body. In contrast to other electron tomography techniques, samples are immobilized in non-crystalline ("vitreous") ice and imaged under cryogenic conditions ($<150^\circ$C ), allowing them to be imaged without dehydration or chemical fixation, which could otherwise disrupt or distort biological structures \cite{b9,b10}. Cryo-electron tomography (Cryo-ET) \cite{b11,b12,b13} enables the 3D visualization of structures at close-to-native state and in sub-molecular resolution within single cells \cite{b14,b15,b16,b17}.

\subsection{Inception3D Network}

\cite{b18} propose a 3D variant of tailored inception network \cite{b19}, denoted as Inception3D. Inception network is a recent successful CNN architecture that has the ability to achieve competitive performance with relatively low computational cost \cite{b18}.
CNN \cite{b17} are well-known for extracting features from a image by using convolutional kernels and pooling layers to emulates the response of an individual to visual stimuli. This work \cite{b20} is the first application of deep learning for systematic structural discovery of macromolecular complexes among large amount (millions) of structurally highly heterogeneous particles captured by Cryo-ET. It represents an important step towards large scale systematic detection of native structures and spatial organizations of large macromolecular complexes inside single cells.

\subsection{Mixup: Data-Dependent Data Augmentation}
Large deep neural networks are powerful, but exhibit undesirable behaviors such as memorization and sensitivity to adversarial examples. \cite{b21} propose mixup, a simple learning principle to alleviate these issues. Essentially, mixup trains a neural network on convex combinations of pairs of examples and their labels. By doing so, mixup regularizes the neural network to favor simple linear behavior in-between training examples, and here is how the mixup training loss is defined:

\begin{math}
\mathcal{L}(\theta) = \mathbb{E}_{x_1,y_1\sim p_{train}} \mathbb{E}_{x_2,y_2\sim p_{train}} \mathbb{E}_{\lambda\sim\beta(0.1)} \ell(\lambda x_1 + (1 - \lambda) x_2, \lambda y_1 + (1 - \lambda) y_2)
\end{math}

\subsection{Focal loss}

\cite{b22,b49} discover that the extreme foreground-background class imbalance encountered during training of dense detectors is the central cause. We propose to address this class imbalance by reshaping the standard cross entropy loss such that it down-weights the loss assigned to well-classified examples. \cite{b23} proposed a novel Focal Loss focuses training on a sparse set of hard examples and prevents the vast number of easy negatives from overwhelming the detector during training.

\subsection{Sampling}

\subsubsection{Oversampling}

Oversampling method achieves sample balanced by increasing the number of minority samples in the classification. The most straightforward way is to simply copy a few samples to form multiple records. However, the disadvantage of this method is that if there are few sample features, it may lead to over-fitting problems. Improved oversampling method by adding random noise, interference data, or certain rules to generate new synthetic samples in a few classes, such as SMOTE algorithm. The process is described about SMOTE method in Algorithm 1.

 \begin{breakablealgorithm}
  \caption{ SMOTE(T, N, K)}
  \label{alg:Framwork}
  \begin{algorithmic}[1]
    \Require
      Number of minority class Samples $T$; 
      
      Amount of SMOTE $\%N$; 
      
      Number of nearest neighbors $k$
    \Ensure
      $(N/100) * T$ synthetic minority class samples
    \State \textit{(* If $N$ is less than $100\%$, randomize the minority class samples as only a random percent of them will be SMOTEd.*)}
    \label{code:fram:extract}
    \If {$N<100$}
            \State Randomize the T minority class samples
            \State $T=(N/100)*T$
            \State $N=100$
    \EndIf
    \label{code:fram:trainbase}
    \State $N=(int)(N/100)$ \textit{(*The amount of SMOTE is assumed to be in integral multiples of 100.*)}
    \label{code:fram:add}
    \State $k$=Number of nearest neighbors
    \label{code:fram:classify}
    \State $numattrs$=Number of attributes
    \label{code:fram:select} 
    \State $Sample$[ ][ ]: array for original minority class samples
    \State $newindex$: keeps a count of number of synthetic samples generated, initialized to 0
    \State $Synthetic$[ ][ ]: array for synthetic samples \textit{(*Compute k nearest neighbors for each minority class sample only.*)}
    \For {$i \gets 1$ to $T$}
        \State Compute $k$ nearest neighbors for i, and save the indices in the $nnarray$ 
        \State Populate($N, i, nnarray$)
    \EndFor \\
    \textit{Populate($N, i, nnarray$) (*Function to generate the synthetic samples.*) }
    \While {$N \neq 0$}
        \State Choose a random number between $1$ and $k$, call it $nn$. \State This step chooses one of the $k$ nearest neighbors of $i$. 
        \For{$attr \gets 1$ to $numattrs$}
            \State Compute: \textit{dif = Sample[nnarray[nn]][attr] - Sample[i][attr]}
            \State Compute: $gap$ = random number between 0 and 1
            \State \textit{Synthetic[newindex][attr] = Sample[i][attr] + gap $\ast$ dif}
        \EndFor
        \State $newindex++$
        \State $N=N-1$
    \EndWhile
    \State \Return \textit{(* End of Populate. *)}
  \end{algorithmic}
\end{breakablealgorithm}

\subsubsection{Undersampling}

Another popular method \cite{b24,b46} that results in having the same number of examples in each class. However, as opposed to oversampling, examples are removed randomly from majority classes until all classes have the same number of examples. While it might not appear intuitive, there is some evidence that in some situations undersampling can be preferable to oversampling \cite{b25}. A significant disadvantage of this method is that it discards a portion of available data. To overcome this shortcoming, some modifications were introduced that more carefully select examples to be removed. E.g. one-sided selection identifies redundant examples close to the boundary between classes \cite{b26}. More general approach than undersampling is data decontamination that can involve relabeling of some examples \cite{b27}.

\section{Our Method}

\subsection{Overview of our model}

\begin{figure}[htbp] 
\centering
\includegraphics[width=0.35\textwidth]{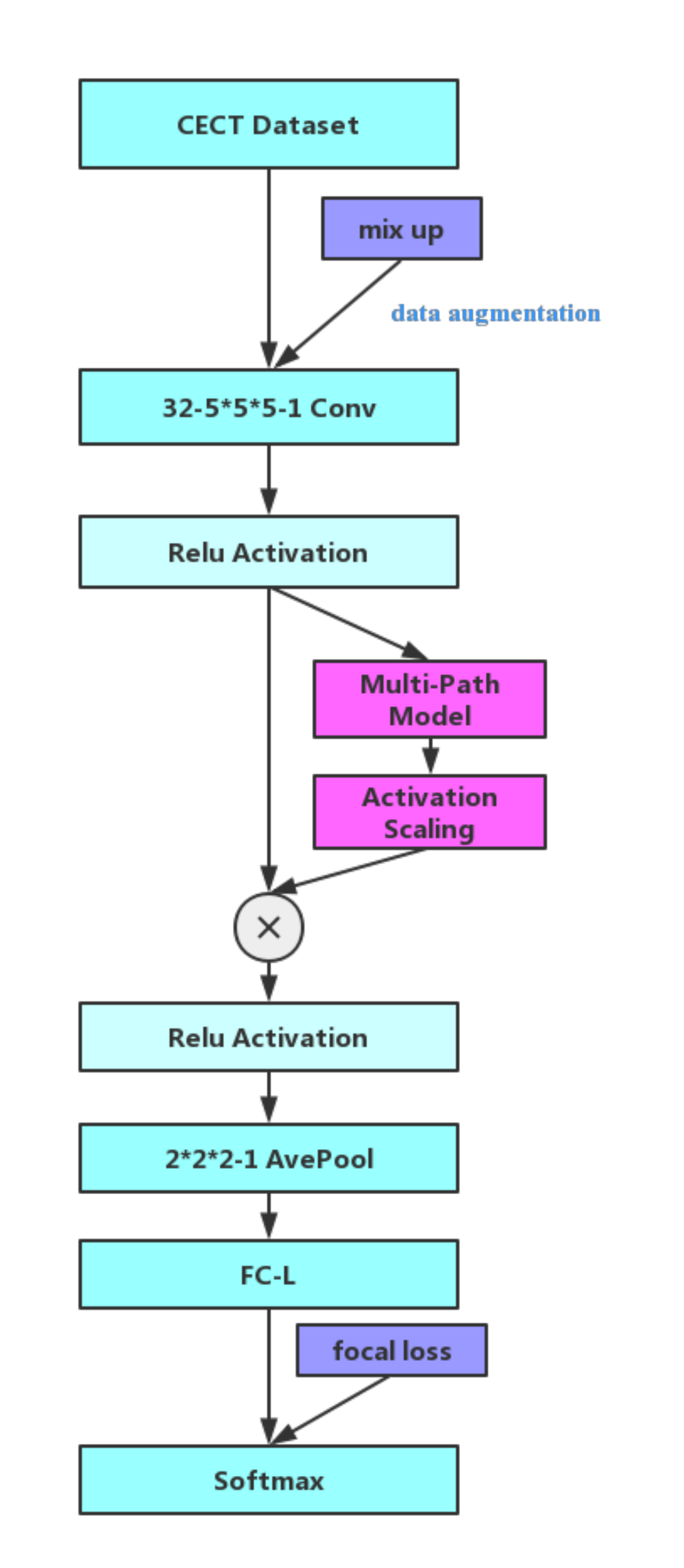} 
\caption{Overview architectures of our model.} 
\label{Fig.main1} 
\end{figure} 

With the inspiration from Inception 3D network \cite{b28,b44,b45}, we propose a novel model termed as multi-path CNN, combaning with mixup and focal loss method, which could significantly improve performance on the imbalanced Cryo-ET data. The overview of our proposed model is as shown in Figure 1.

\subsection{Mixup on the Cryo-ET dataset}

Motivated by \cite{b29,b42,b43}, we introduce a simple and data-agnostic data augmentation routine, termed as mixup. In a nutshell, mixup constructs virtual training examples
\[{\tilde{x}}= \lambda x_i+(1-\lambda )x_j\]
\[{\tilde{y}}= \lambda y_i+(1-\lambda )y_j\] 

\noindent where $x_i$,$x_j$ are raw input vectors, and $y_i$ , $y_j$ are one-hot label encodings
$(x_i,y_i)$ and $(x_j,y_j)$ are two examples drawn at random from our training data, and $\lambda \in [0,1]$. Therefore, mixup extends the training distribution by incorporating the prior knowledge that linear interpolations of feature vectors should lead to linear interpolations of the associated targets.

\subsection{Multi-path CNN}
Many subtomogram data from Cryo-ET are imbalanced due to their different ratio in the cell. However, there is little work that had been done to solve the problem of imbalanced data from Cryo-ET. In this section, we describe detaily our proposed multi-path CNN model showed in Figure 2,  which will be useful to cope with the imbalanced data from Cryo-ET with the combination of the related works from Section 2.

\begin{figure}[htbp] 
\centering
\includegraphics[width=0.5\textwidth]{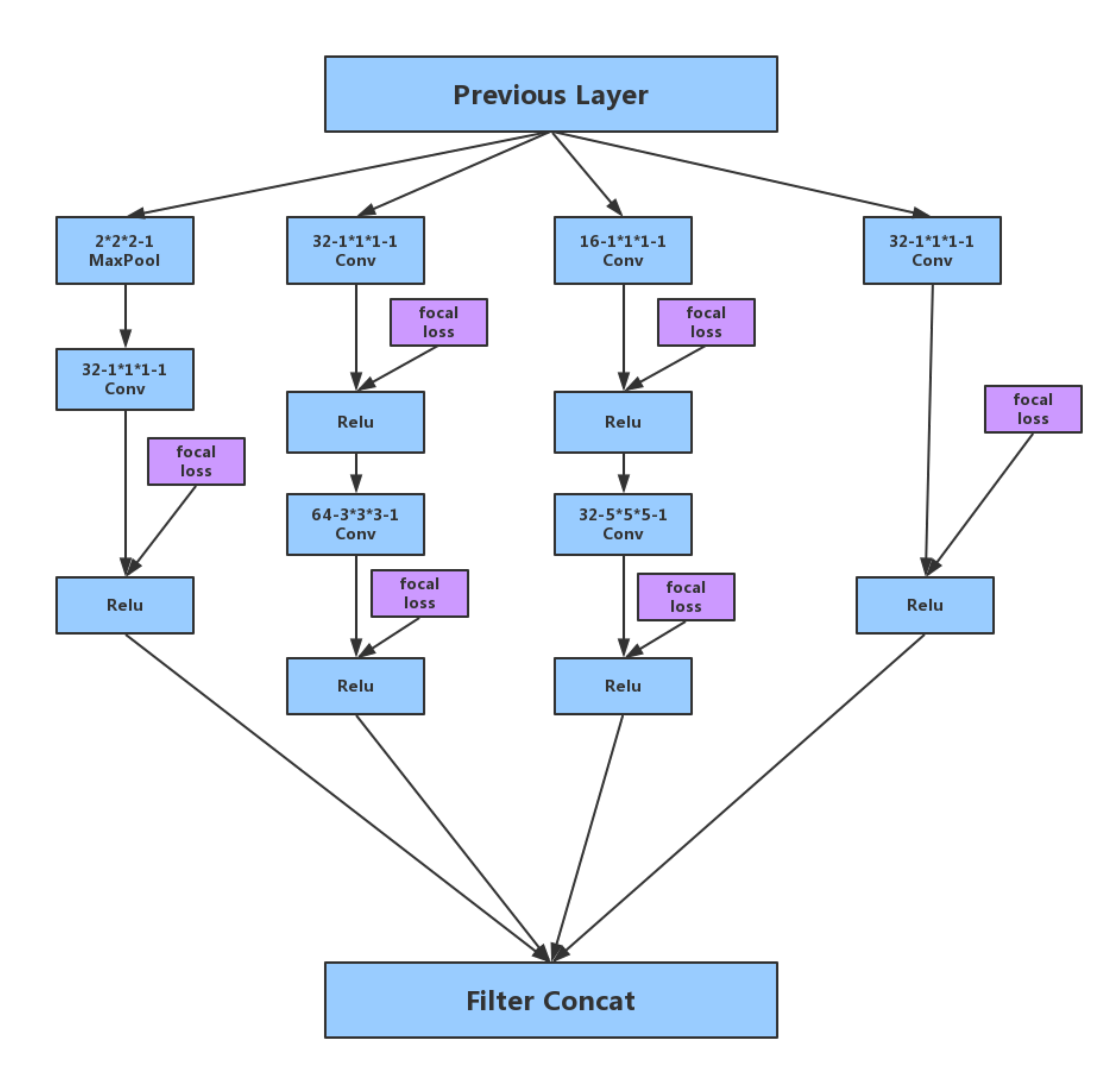}
\caption{Model of multi-path CNN} 
\label{Fig.main2} 
\end{figure} 

Unlike traditional CNN which only has a single path with serial combinations of convolutional kernels and pooling layers, our multi-path CNN has multiple parallel combinations of convolutional kernels and pooling layers, based on the composition of our imbalanced data. More specifically, the number of classes from Cryo-ET equals to the number of parallel paths in our multi-path CNN model. Each path will try to learn from the imbalanced Cryo-ET data and become the best classifier corresponding to a certain type of data before the concatenate layer. The reason about firstly trying to determine each single path for each type of data from Cryo-ET is that it will be easier to find a classifier through deep learning that will behave well in recognizing a single class of data, regardless of other classes.

In order to find the best classifier for a certain type of data from Cryo-ET, we firstly carry out a lot of experiments with serial CNN and find its best structure to recognize a single type from the imbalanced dataset. All paths will be concatenated together when they are identified respectively. It is worth mentioning that sampling methods have not been used in the whole process since it may lead to overfitting or underfitting. Each single path learns how to do the classification job from the original imbalanced data and whole multi-path CNN learns how to balance between these paths and make a more precise decision towards all the types of data from Cryo-ET.

\subsection{Filter Concat}

The final model obtains the most suitable convolution kernel on each path, so that the model effect is optimal, and then combines the best parameters learned by the models on the four paths, and finally enters the new pooling layer through a filter. The final classification result is obtained by softmax through the full connection of the L layer.

\subsection{Focal Loss for imbalanced classification}
We use an $\alpha$-balanced variant of the focal loss:
\[FL(p_t) = -\alpha _t(1-p_t)^{\gamma }log(p_t)\]
We adopt this form in our experiments as it yields slightly improved accuracy over the non-$\alpha$-balanced form. Finally, we note that the implementation of the loss layer combines the sigmoid operation for computing p with the loss computation, resulting in greater numerical stability.

\begin{table*}[ht]
\centering

\setlength{\tabcolsep}{10mm}{
\begin{tabular}{l|c|c}
\hline
               & \multicolumn{1}{l|}{pridicted positives} & \multicolumn{1}{l}{predicted negatives} \\ \hline
Real positives & TP                                       & FN                                      \\
Real negatives & FP                                       & TN                                      \\ \hline
\end{tabular}}
\caption{Confusion matrix for binary classification}
\end{table*}

\begin{table*}[]
\centering
\setlength{\tabcolsep}{6mm}{
\begin{tabular}{c|cccc}
\hline
\multicolumn{1}{l|}{} & proteasome\_d & ribosome & TRiC  & proteasome\_s \\ \hline
Number                & 1043          & 80       & 125   & 386           \\ \hline
Ratio(\%)             & 63.481        & 4.869    & 7.608 & 23.494        \\ \hline
\end{tabular}}
\caption{Class ratio of Cryo-ET dataset.}
\end{table*}

\subsection{Evaluation metrics}
Evaluation metrics play an important role in assessing the classification performance and guiding the model design. Most of the traditional methods dealing with the imbalanced data concentrate on binary classification. In binary classification problem, class labels can be divided as positive and negative. As the confusion matrix shows in Table 1, true positive (TP) and true negative (TN) denote the number of positive and negative samples that are correctly classified while false negative (FN) and false positive (FP) denote the number of positive and negative samples that are wrongly classified.

\begin{figure*}
\centering
\subfigure[proteasome\_d]{\label{fig:subfig:a}
\includegraphics[width=0.25\linewidth]{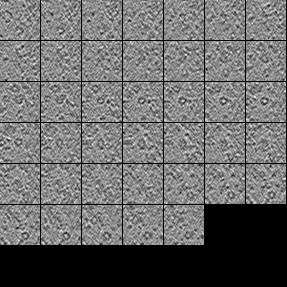}}
\hspace{0.01\linewidth}
\subfigure[ribosome]{\label{fig:subfig:b}
\includegraphics[width=0.25\linewidth]{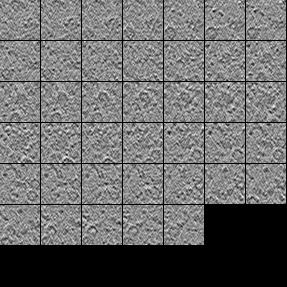}}
\vfill
\subfigure[TRiC]{\label{fig:subfig:a}
\includegraphics[width=0.25\linewidth]{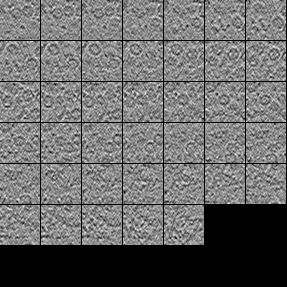}}
\hspace{0.01\linewidth}
\subfigure[proteasome\_s]{\label{fig:subfig:b}
\includegraphics[width=0.25\linewidth]{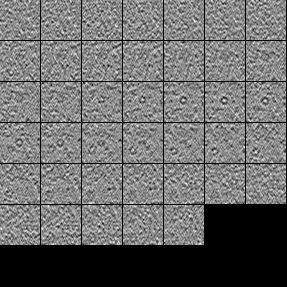}}
\caption{The 2D visualizations of the 3D macro molecules}
\label{fig:subfig}
\end{figure*}

Accuracy is the most commonly used metric to evaluate model performance, however, it is no longer a proper measure in imbalanced classification problem since the minor class has minimal impact on accuracy compared with the major class. To solve the problem, a pair of metrics, precision and recall, have been adopted.

\[precision = \frac{{TP}}{{TP + FP}}\]

\[recall = \frac{{TP}}{{TP + FN}}\]

Meanwhile, F-Score is used to integrate precision and recall into a single metric for convenient evaluation of model.

\[{{\rm{F}}_\beta }{\rm{ = }}(1 + {\beta ^2}) \cdot \frac{{precision \cdot recall}}{{({\beta ^2} \cdot precision) + recall}}\]

Where $\beta$ represents the weight between precision and recall. During our evaluation process, we set $\beta$ = 1 since we regard precision and recall has the same weight thus $F_1$-score is adopted.

However, in multi-class classification, we use Macro $F_1$-Score to evaluate the result.
\[Macro{\kern 1pt} {\kern 1pt} {\kern 1pt} {\kern 1pt} {F_1}{\rm{ = }}\frac{{\sum\limits_{n = 1}^N {{F_{1n}}} }}{N}\]
where n represents the number of classification and $F_{1n}$ is the $F_1$ score on nth category.

G-mean is another recognized metric derived from confusion matrix.

\[G - mean = \sqrt {\frac{{TP}}{{TP + FN}} + \frac{{TN}}{{TN + FP}}} \]

In multi-class classification, we also use Macro G-mean to evaluate the result, measure the balanced performance of a learning model.

\[Macro{\kern 1pt} {\kern 1pt} {\kern 1pt} {\kern 1pt} {G-mean}{\rm{ = }}\frac{{\sum\limits_{n=1}^N {{G_{n}}} }}{N}\]
where n represents the number of classification and $G_{n}$ is the G-mean on nth category.

\section{Experiments}
\subsection{Dataset details}
Furthermore, reference-free classification and averaging were tested on a dataset consisting of 125 TCP-1 ring complex (TRiC) subtomograms, 386 single  capped proteasome (proteasome\_s) subtomograms, 1043 double capped proteasome (proteasome\_d) subtomograms, and 80 ribosome subtomograms extracted from a tomogram of rat neuron with expression of poly-GA aggregate. All subtomorgams were two times binned to size $40^3$ (voxel size: 1.368 nm). The tilt angle range was $- 50^\circ$ to $+ 70^\circ$.

The four types of macro molecules in our Cryo-ET dataset, which are proteasome\_d, ribosome, TRiC, proteasome\_s, whose imbalanced ratio are shown in Table 2. The Figure 3 are the 2D visualizations of the 3D macro molecules.

% \usepackage{multirow}
\begin{table*}[]
\centering
\setlength{\tabcolsep}{2.5mm}{
\begin{tabular}{c|cc|cc|cc|cc}
\hline
\multirow{2}{*}{Classes in Cryo-ET} & \multicolumn{2}{c|}{path 1}   & \multicolumn{2}{c|}{path 2}   & \multicolumn{2}{c|}{path 3}   & \multicolumn{2}{c}{path 4}    \\ \cline{2-9} 
                                 & F1            & G-mean        & F1            & G-mean        & F1            & G-mean        & F1            & G-mean        \\ \hline
proteasome\_d                    & \textbf{71.7} & \textbf{81.2} & 68.3          & 79.2          & 69.0          & 78.5          & 70.8          & 80.2          \\
ribosome                         & 69.7          & 78.1          & \textbf{74.2} & \textbf{82.6} & 72.3          & 79.4          & 71.4          & 77.6          \\
TRiC                             & 72.6          & 81.2          & 71.7          & 79.8          & \textbf{74.8} & \textbf{83.8} & 74.1          & 82.4          \\
proteasome\_s                    & 71.2          & 80.4          & 69.2          & 78.5          & 72.7          & 81.9          & \textbf{73.3} & \textbf{83.0} \\ \hline
\end{tabular}}
\caption{Binary classification by four different paths}
\end{table*}

% \usepackage{multirow}
\begin{table*}[]
\centering
\setlength{\tabcolsep}{7mm}{
\begin{tabular}{c|c|c|l}
\hline
\multirow{2}{*}{Model}                                     & \multicolumn{3}{c}{Cryo-ET}                            \\ \cline{2-4} 
                                                           & Macro F1        & \multicolumn{2}{c}{Macro G-mean}  \\ \hline
Multi-path CNN                                             & 68.1            & \multicolumn{2}{c}{78.1}          \\
Multi-path CNN with boosting                               & 69.3            & \multicolumn{2}{c}{80.3}          \\
Multi-path CNN with bagging                                & 69.0            & \multicolumn{2}{c}{79.8}          \\
Multi-path CNN with SMOTE                                  & 69.7            & \multicolumn{2}{c}{78.6}          \\
Multi-path CNN with undersampling                          & 69.1            & \multicolumn{2}{c}{77.3}          \\
Multi-path CNN with GP                         & 70.4            & \multicolumn{2}{c}{78.2}          \\
Multi-path CNN with mixup                         & 70.9            & \multicolumn{2}{c}{80.2}          \\
Multi-path CNN with focal loss                         & 71.1            & \multicolumn{2}{c}{80.7}          \\
Multi-path CNN with SMOTE + boosting                         & 70.1            & \multicolumn{2}{c}{78.2}          \\
Multi-path CNN with SMOTE + bagging                         & 71.4            & \multicolumn{2}{c}{80.2}          \\
Multi-path CNN with SMOTE + GP                         & 72.6            & \multicolumn{2}{c}{81.5}          \\
\multicolumn{1}{l|}{\textbf{Multi-path CNN with mixup+focal loss}} & \textbf{73.6}   & \multicolumn{2}{c}{\textbf{84.7}} \\ \hline
\end{tabular}}
\caption{Multi-path CNN combined with different strategies on imbalanced Cryo-ET}
\end{table*}

In order to train and test our multi-path CNN, we shuffle and split our dataset with two parts. There are 1307 samples in the training set and 327 samples in the testing set.

\subsection{Baseline Methods}

\subsubsection{Bagging}

\cite{b30} introduced the concept of bootstrap
aggregating to construct ensembles. It consists in
training different classifiers with bootstrapped replicas of
the original training data-set. That is, a new data-set is
formed to train each classifier by randomly drawing (with
replacement) instances from the original data-set (usually,
maintaining the original data-set size). Hence, diversity
is obtained with the resampling procedure by the usage
of different data subsets. Finally, when an unknown instance
is presented to each individual classifier, a majority
or weighted vote is used to infer the class \cite{b31}.

\subsubsection{Boosting}

Boosting (also known as ARCing, adaptive resampling
and combining) was introduced by Schapire in
1990 \cite{b32,b33,b34}. Schapire proved that a weak learner (which is
slightly better than random guessing) can be turned into a
strong learner in the sense of probably approximately correct
(PAC) learning framework. AdaBoost \cite{b35} is the most
representative algorithm in this family, it was the first applicable
approach of Boosting, and it has been appointed as
one of the top ten data mining algorithms \cite{b36}.

% \usepackage{multirow}
\begin{table*}[]
\centering
\setlength{\tabcolsep}{5.5mm}{
\begin{tabular}{c|c|c}
\hline
\multirow{2}{*}{Model}                & \multicolumn{2}{c}{Cryo-ET}     \\ \cline{2-3} 
                                      & Macro F1      & Macro G-mean  \\ \hline
\textbf{Multi-path CNN with mixup + focal loss} & \textbf{73.6} & \textbf{84.7} \\ \hline
GSVM-RU                               & 69.5          & 80.3          \\ \hline
\end{tabular}}
\caption{Comparison between multi-path CNN model and traditional method}
\end{table*}

% \usepackage{multirow}
\begin{table*}[]
\centering
\setlength{\tabcolsep}{5.5mm}{
\begin{tabular}{c|c|c}
\hline
\multirow{2}{*}{Model}               & \multicolumn{2}{c}{Cryo-ET}     \\ \cline{2-3} 
                                     & Macro F1      & Macro G-mean  \\ \hline
\textbf{Four-path CNN with mixup + focal loss} & \textbf{73.6} & \textbf{84.7} \\
GSVM-RU(based on four classes)                & 69.5          & 80.3          \\ \hline
Three-path CNN with mixup + focal loss         & 74.3          & 85.5          \\
GSVM-RU(based on three classes)               & 70.1          & 81.6          \\ \hline
Two-path CNN with mixup + focal loss           & 76.4          & 87.2          \\
GSVM-RU(based on two classes)                 & 71.2          & 81.9          \\ \hline
\end{tabular}}
\caption{Comparison of different number of paths in multi-path CNN}
\end{table*}

\subsubsection{Genetic Programming (GP)}

GP\cite{b37} is an evolutionary algorithm technique inspired from biological evolution to find computer programs that perform a user-defined task, which can evolve biased classifiers when data sets are unbalanced. In GP, programs representing different solutions to a problem are combined with other programs to create new hopefully better programs; this process is repeated over a number of generations until a good solution is evolved \cite{b38,b39,b40}. \cite{b41} proposed GP methods utilize the unbalanced data “as is” in the learning phase, requiring no prior knowledge about the problem domain, to evolve classifiers with good classification ability on both minority and majority classes.

\subsection{Identification of each path in Multi-path CNN}

Before carrying out the experiment corresponding to the whole model in Figure 2, we have carried out lots of experiments to identify each suitable path in the multi-path CNN model. We name the four paths with path 1, path 2, path 3 and path 4, from left to right in the model in Figure 2. Each path has the best binary classification result on one of the four classes in Cryo-ET. For example, as shown in Table 3, path 1 behaves best on proteasome\_d, while path 3 has the best result on TRiC class. During the experiment in each single path, we degenerate the multi-class classification problem with binary classification, say, the CNN branch in path 1 will only tell whether the input belongs to proteasome\_d class and path 3 will only judge whether the input belongs to TRiC class.

\begin{figure*}[htbp] 
\centering
\includegraphics[width=0.6\textwidth]{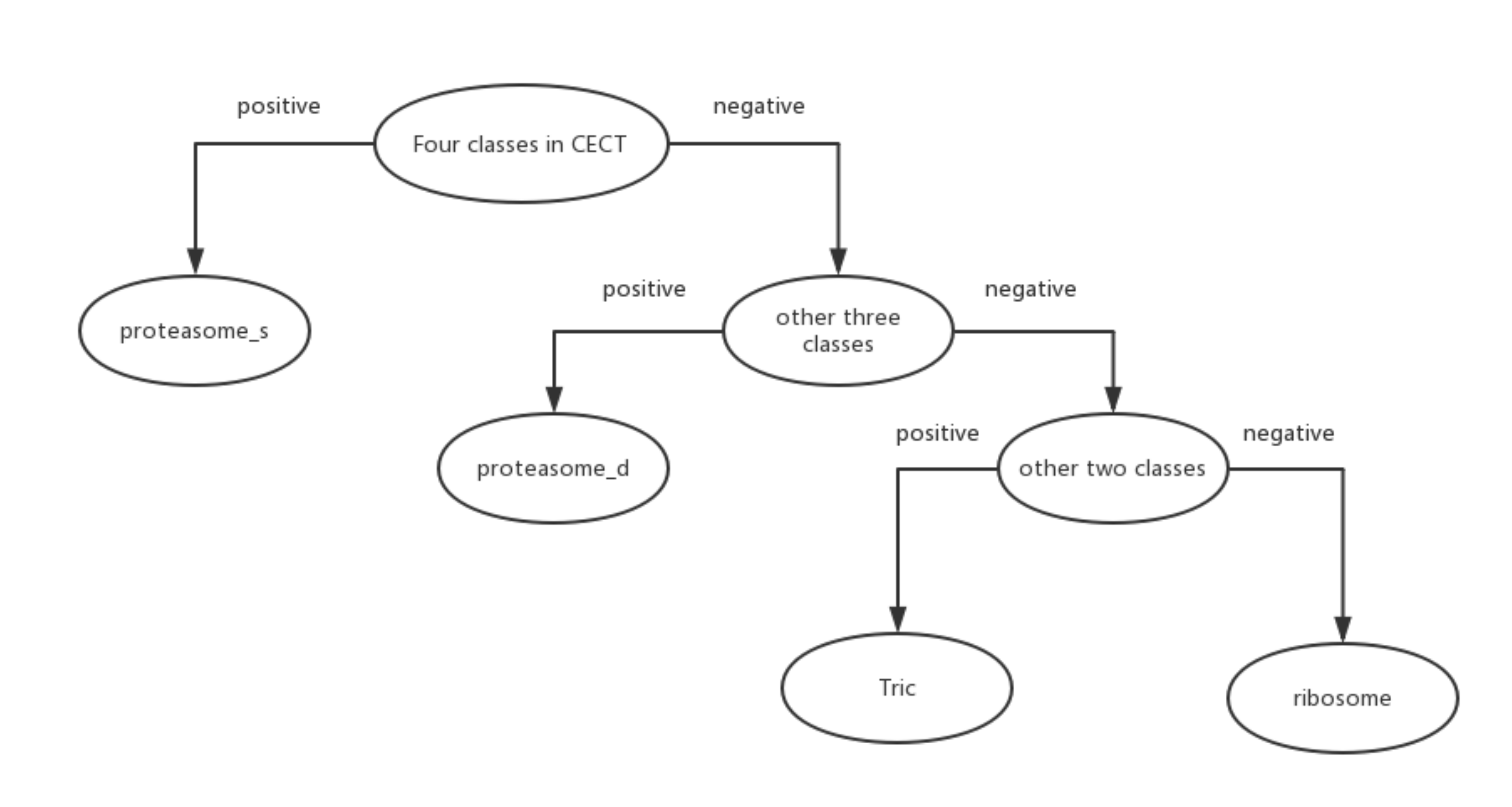} 
\caption{Multiclass classification achieved by binary classifier} 
\label{Fig.main1} 
\end{figure*}

\subsection{Multi-path CNN with traditional strategies on imbalanced dataset}

We have done some experiments to combine the multi-path CNN with recognized and effective strategies like boosting, bagging, SMOTE, focal loss and mixup method, towards the imbalanced dataset. The results are shown in Table 4.

Experiments have also been done to compare the multi-path CNN with traditional classifier, Granular Support Vector Machine with Repetitive Undersampling(GSVM-RU), \cite{b47,b48}. Besides SVM modeling, GSVM-RU adds another hyper-parameter G, the number of negative granules. To solve the problems in multi-class classification with binary classifier, we use the method in Figure 4 to decompose the multi-class problem into a binary class problem. The experiment results are shown in Table 5.

Several pairs of experiments have also been done to show that our model be adjusted to two or three classes classification problem, rather than restricted in four classes classification problem. We conduct GSVM-RU on different number of classes as baseline for comparison. The results are shown in the Table 6.

\subsection{Discussion}

Cryo-ET has become a powerful tool for 3D visualization of cellular components in sub-molecular resolution and near-primary ecology \cite{b50}. However, imbalanced classification in cellular tomograms is difficult due to the high complexity of image content and imaging limitations. In order to complement the existing method, in this paper, we propose a  multi-path CNN combined with mixup and focal loss strategy which will have the best classification result on the imbalanced data from Cryo-ET. The above experiment results demonstrate the power of our approach and they have also indicated that by changing the number of paths in our multi-path, the model can be adapted to cope with imbalanced classification problems with different number of classes. The work provides useful steps for imbalanced classification in cell tomography. To the best of our knowledge, our work is the first application of CNN-based network with focal loss and mixup method in Cryo-ET data analysis. Our approach is a useful complement to current technology

\section{Conclusion}
In this paper, we apply the method of dealing with imbalanced data to the classification of cell macromolecular complexes for the first time, which opened up a new path for cell classification in the field of computational biology. In order to solve the imbalanced data problem from Cryo-ET, we propose a multi-path CNN model combined with recognized strategies to deal with data imbalance issue like sampling, bagging and boosting and genetic programming. We have also made combinations among the methods and with our model. The multi-path CNN model consists of several independent paths that behave best in each class respectively. By adjusting the number of the paths in the model, we can deal with a more generalized classification problem with different number of classes. Experiments and comparisons with traditional classifiers have shown that the model can work effectively on the imbalanced data from Cryo-ET. In the future, we will also consider more issues in the field of computer and bio-related technologies to promote the development of computational biology.

\section{Acknowledgement}

This work was supported in part by U.S. National Institutes of Health (NIH) grant P41 GM103712.

% \section*{References}

% Please number citations consecutively within brackets \cite{b1}. The 
% sentence punctuation follows the bracket \cite{b2}. Refer simply to the reference 
% number, as in \cite{b3}---do not use ``Ref. \cite{b3}'' or ``reference \cite{b3}'' except at 
% the beginning of a sentence: ``Reference \cite{b3} was the first $\ldots$''

% Number footnotes separately in superscripts. Place the actual footnote at 
% the bottom of the column in which it was cited. Do not put footnotes in the 
% abstract or reference list. Use letters for table footnotes.

% Unless there are six authors or more give all authors' names; do not use 
% ``et al.''. Papers that have not been published, even if they have been 
% submitted for publication, should be cited as ``unpublished'' \cite{b4}. Papers 
% that have been accepted for publication should be cited as ``in press'' \cite{b5}. 
% Capitalize only the first word in a paper title, except for proper nouns and 
% element symbols.

% For papers published in translation journals, please give the English 
% citation first, followed by the original foreign-language citation \cite{b6}.

\vspace{12pt}
% \color{red}
% IEEE conference templates contain guidance text for composing and formatting conference papers. Please ensure that all template text is removed from your conference paper prior to submission to the conference. Failure to remove the template text from your paper may result in your paper not being published.

\end{document}